\documentstyle[preprint,aps]{revtex}
\global\firstfigfalse
\global\firsttabfalse

\begin{document}
\draft
\title{Enhancement of the tunneling density of states in
Tomonaga--Luttinger liquids}
\author{Yuval Oreg and Alexander M. Finkel'stein \cite{Lan}}
\address{Department of Condensed Matter Physics\\
The Weizmann Institute of Science, Rehovot ISRAEL 76100}
\date{\today }
\maketitle

\begin{abstract}
We have calculated the tunneling density of states (DOS) at the location of
a backward scattering defect for quantum wires and for edge state electrons
in quantum Hall systems. A singular enhancement of the DOS arises as a
result of the combined effect of multiple backward scattering together with
a repulsive electron---electron interaction.
\end{abstract}

\pacs{PACS:73.40.Gk, 72.15.Fk, 73.40.Hm}

With the rapid advance of the submicron technology\cite{EMS:Kastner92} the
fabrication of one--dimensional (1--d) quantum wires has become reality. The
properties of these wires are expected to be unusual. It is known that the
electron--electron interaction in a 1--d electron gas, when away from the
density--wave or from the superconductivity instabilities, leads to the
Tomonaga--Luttinger (TL) liquid behavior\cite{EE1D:Mattis65}. The
spectacular feature of the TL--liquid is the vanishing of the
single--particle DOS at the Fermi energy\cite
{EE1D:Dzyaloshinskii74,EE1D:Luther74}. In this work we calculate the DOS of
the TL--liquid at the location of a defect center that causes backward
scattering of the conduction electrons. By mapping the problem onto a
Coulomb gas theory we show that the DOS diverges at energies close to the
Fermi energy when the electron--electron interaction is repulsive and not
too strong; i.e., the tunneling DOS in the vicinity of a backward scatterer
is in clear contrast with the DOS in a clean TL--liquid or far away from the
scattering center. It has already been noted by us in connection with the
study of the Fermi--edge singularity in 1--d that the low energy physics of
the backward scattering together with the electron--electron repulsion
resembles the physics of the Kondo resonance\cite{XR:Oreg95}. The singular
enhancement of the DOS is a consequence of a many body effect of a similar
type.

Recently, considerable efforts have been directed towards the study of the
transport properties of the 1--d TL--liquids\cite
{EE1D:Kane92a,EE1D:Furusaki93,EE1D:Matveev93,EE1D:Ogata94,EE1D:Fendley95,EE1D:Tarucha95,FQHE:Milliken94,FQHE:Moon93}.
For a repulsive electron--electron interaction it has been predicted that
at zero temperature even a single weak backward scatterer eventually
causes the conductance to vanish. It is widely accepted that the low
energy physics of this system can be described by two semi--infinite lines
connected by a weak link junction (e.g., in Ref.\cite{EE1D:Kane92a} the
vanishing of the conductance has been traced to the fact that the tunneling
DOS into the end of a semi--infinite TL--liquid vanishes at the Fermi
energy). However, as found in the present work, in the vicinity of the
backward scatterer the DOS is enhanced for repulsive electron--electron
interaction. We believe that the description of the low energy physics of
this problem by two disconnected wires should be exploited with caution.

Besides the quantum wires the TL--liquid behavior can be displayed by edge
state electrons in quantum Hall devices\cite
{FQHE:Milliken94,FQHE:Moon93,FQHE:Wentot,QHE:Oreg95}. The tunneling DOS in
the quantum Hall systems will be discussed at the end of the paper.

The renormalization group treatment shows that the effective amplitude of
the backward scattering increases and becomes strong when the energy of the
scattered electrons approaches the Fermi energy\cite
{EE1D:Kane92a,EE1D:Mattis74b}. Therefore, at low temperatures one needs to
understand the physics of the strong coupling regime. From the experience of
the study of local defect problems in metals it is known that mapping the
problem onto a Coulomb gas theory can be instructive (see e.g., Refs.\cite
{KE&AMM:Anderson70a}). The discussed problem has been mapped onto a theory
of a neutral gas of positively and negatively charged classical particles
interacting via a logarithmic potential\cite{EE1D:Kane92a}. These charges
are located on a line and they describe the time history of the
backscattering events. Unlike the Kondo problem, this problem is described
by a non alternating Coulomb gas. The physics of this gas have been well
studied. There are two phases separated at a critical temperature, $T_{cr}$,
by a transition of the Kosterlitz--Thouless type\cite
{RFS:Kosterlitz73,EE1D:Bulgadaev82}. The temperature of the Coulomb gas,
$T_{gas}$, is determined by the electron--electron interaction of the
original problem. At low temperatures, $T_{gas}<T_{cr}$, the particles form
dipoles, while in the hot phase, $T_{gas}>T_{cr}$, the dipoles dissociate
and the gas is in the plasma state. From the renormalization group analysis
it follows that when the electron--electron interaction is repulsive the
system is in the hot plasma phase, while the dipole phase corresponds to the
attractive electron--electron interaction. In the plasma phase the
logarithmic interactions between charged particles are screened--off at
distances exceeding the radius of screening $\tau _{scr}$. To describe the
strong coupling regime of the backward scattering problem in the TL--liquid
we will utilize the physics of screening in the plasma phase of the Coulomb
gas.

For simplicity, we start with the spinless case and will include the spin
degrees of freedom later. The Hamiltonian of the TL--liquid in 1--d can be
written in terms of the bosonic field operators $\phi $ and $\tilde{\phi}$ as
\cite{EE1D:Solyom79,EE1D:Emery79}
\begin{equation}
H_{TL}^{}=\frac{v_F}{2g}\int dx\left( \left( \frac{d\phi }{dx}\right)
^2+\left( \frac{d\tilde{\phi}}{dx}\right) ^2\right) ,  \label{eq:htl}
\end{equation}
where $g=\sqrt{\frac{1-\gamma }{1+\gamma }},$ $\gamma =\frac V{\left( 2\pi
v_F+V\right) },$ $v_F$ is the Fermi velocity and $V$ describes the
density--density interaction with momentum transfers smaller than the Fermi
momentum $k_F$. In Eq.~(\ref{eq:htl}) the operator $\frac{d\phi (x)}{dx}$ is
proportional to the deviation of the electron density from its average
value, and $\frac{d\widetilde{\phi }(x)}{dx}$ is proportional to the current
density; $\phi $ and its dual partner $\tilde{\phi}$ are conjugate
variables, i.e., $\left[ \frac{d\phi (x)}{dx},\tilde{\phi}(y)\right] =i{\bf
\delta }(x-y)$. Hamiltonian (\ref{eq:htl}) describes the 1--d electron
liquid when the backward scattering can be ignored in the
processes of the electron--electron interaction.
 The field operators $\psi _{R(L)}(x)$ of
electrons with momenta close to $\pm k_F$ can be represented using the
bosonization
technique\cite{EE1D:Luther74,EE1D:Mattis74b,EE1D:Solyom79,EE1D:Emery79}
as \label{eq:fermion} \begin{equation}
\psi _{R(L)}(x)=\frac{e^{\pm ik_Fx}}{\sqrt{2\pi \eta }}\exp {\left[ -\frac i2
\left( \frac{4\pi }\beta \tilde{\phi}\pm \beta \phi \right) \right] ,}
\label{eq:f-b}
\end{equation}
where $\beta ^2=4\pi g$ and $\eta ^{-1}$ is an ultraviolet cutoff of the
order of the conduction band width.

 We concentrate below on the $2k_F$--backward scattering only.
 It will be argued later that the DOS is not
influenced by the forward scattering. The backward scattering induced by a
defect located at the point $x=0$ is described by a term
\begin{equation}
{\qquad H}_{bs}={U(2k}_F{)\psi }_R^{\dagger }(0){\psi }_L(0)+{U}^{*}{(2k}_F{
)\psi }_L^{\dagger }(0){\psi }_R(0),  \label{eq:scatt}
\end{equation}
where $U(2k_F)$ is the $2k_F$--Fourier transform amplitude of the
scattering potential, and ${\psi }_{R(L)}(0)={\psi }_{R(L)}(x=0)$. In the
bosonic representation ${H}_{bs}$ can be written as
\begin{equation}
H_{bs}=-\frac{\delta _{-}}\pi \frac{v_F}\eta \cos \left( \beta \phi
(0)+\varphi _u\right) ,  \label{eq:bsc}
\end{equation}
where $\delta _{-}=\left| {U(2k}_F{)}\right| /v_F$ and ${U(2k}_F{)=-}\left| {
U(2k}_F{)}\right| e^{i\varphi _u}$.

The local tunneling DOS will be found as
\begin{equation}
{\qquad {\varrho }(\epsilon ,x)=-}\frac 1\pi \text{Im}\left\{ \int_0^{\infty
}G(\tau ;x,x)e^{i\epsilon _n\tau }d\tau \right\} _{i\epsilon _n\rightarrow
\epsilon +i\delta }\text{,}  \label{eq:tDOS}
\end{equation}
where $G(\tau ;x,x)=-\left\langle {\cal T}_\tau \psi (\tau ,x)\psi ^{\dagger
}(0,x)\right\rangle $ is the Matsubara Green's function of the electrons at
the point $x$ and $\psi (\tau ,x)=\psi _L(\tau ,x)+\psi _R(\tau ,x)$. In the
absence of scattering $G(\tau ;x,x)$ can be readily obtained using
representation (\ref{eq:f-b}) and the fact that Hamiltonian (\ref{eq:htl})
is quadratic. This leads to $G(\tau ;x,x)\sim \exp -\frac 12\left( g{\cal D}
_\phi +g^{-1}{\cal D}_{\widetilde{\phi }}\right) ,$ where ${\cal D}_\phi
(\tau )=2\pi \left\langle \phi (\tau ,x)\phi (0,x)-\phi (0,x)^2\right\rangle
=\log \left( 1+\frac{v_F\tau }{\eta g}\right) $ is the Green function of the
$\phi $--operators, and in a similar way ${\cal D}_{\tilde{\phi}}(\tau
)=\log \left( 1+\frac{v_F\tau }{\eta g}\right) $. As a result\cite
{EE1D:Dzyaloshinskii74,EE1D:Luther74},
\begin{equation}
{{\varrho }(\epsilon ,x)}\sim \epsilon ^{\left( g-1\right) ^2/2g}.
\label{eq:fldos}
\end{equation}
To study the DOS at the location of the backward scatterer, $x=0$, we treat $
G(\tau ;0,0)$ in the interaction representation with respect to $H_{bs}$.
When the backward scattering is written as in Eq.~(\ref{eq:bsc}), each term
in the perturbation series for $G(\tau ;0,0)$ can be calculated using
the Baker--Ha\"{u}sdorf formula repeatedly. This procedure gives
straightforwardly a representation of the Green function in terms of
partition functions of a one--dimensional Coulomb gas of classical particles:
\begin{equation}
G(\tau ;0,0)\sim \tau ^{-\frac 1{2g}}\left( Z_e(\tau )-Z_o(\tau )\right) /Z.
\label{eq:gcg}
\end{equation}
The factor $\tau ^{-\frac 1{2g}}$ originates from the $\tilde{\phi}$--field
factors in the bosonic representation of the fermion--field operators (see
Eq.~(\ref{eq:f-b})). Since the operators $\psi (\tau ,x)$ have left and right
components, $Z_e(\tau )$ and $Z_o(\tau )$ contain four contributions each,
e.g., $Z_e(\tau )=Z_e^{+-}(\tau )+Z_e^{-+}(\tau )+Z_e^{++}(\tau
)+Z_e^{--}(\tau )$. The term $Z_e^{+-}$ is the grand partition function of a
neutral Coulomb gas that has a charge $+\frac 12$ at the point $0$, a
charge $-
\frac 12$ at the point $\tau $, and  an even number of charges $\pm 1$
between them. The other three terms $Z_e^{aa^{\prime }}(\tau )$ are defined
in a similar way, namely the upper indices correspond to the signs of the $\
\pm \frac 12$ charges located at the points $0$ and $\tau $. These
half--charges originate from the $\phi $--field factors of the operators $
\psi _{L(R)}$ in the Green function. $Z_o(\tau )$ is analogous to $Z_e(\tau)$,
but with an odd number of $\pm 1$ charges inside the interval $(0,\tau )$.
 The term in the denominator, $Z$, is the grand partition function of the
Coulomb gas without the additional half--charges. The minus sign in front of
$Z_o(\tau )$ in Eq.~(\ref{eq:gcg}) appears because of the anticommutation of
the fermion operators.  In the discussed Coulomb gases the particles
interact via a logarithmic potential $v\left( \tau -\tau ^{\prime }\right)
=\log \left( 1+\frac{v_F|\tau -\tau^{\prime }|}{\eta g}\right) $, the
fugacity is $g\frac{\delta _{-}}{2\pi }$ and the effective temperature $
T_{gas}=\frac 1{2g}$. Thus, the calculation of the DOS is reduced to the
calculation of correlation functions in the Coulomb gas theory.

To analyze the functions $Z_e^{aa^{\prime }}(\tau )$ and $Z_o^{aa^{\prime
}}(\tau )$ we integrate out the field $\phi (t,x)$ in the entire space
except the point of the backward scatterer location and reformulate the
problem in terms of a functional integral over $\phi (t)\equiv \phi (t,
x=0)$. The difference $\Delta Z_{}^{aa^{\prime }}=Z_e^{aa^{\prime }}(\tau
)-Z_o^{aa^{\prime }}(\tau )$ can be obtained using the effective action
\begin{equation} S^{aa^{\prime }}(\phi )=\frac 12\int dtdt^{\prime }\phi
(t)v^{-1}(t-t^{\prime })\phi (t^{\prime })+\frac{2\delta _{-}}{\pi \eta
}\int dt\cos \left( \beta \phi (t)+W(t)\right) +a\frac i2\beta \phi
(0)+a^{\prime }\frac i2\beta \phi (\tau ).\label{eq:action} \end{equation}
Here $v^{-1}$ is the inverse of the potential $v$; the potential $W(t)=\pi
\left( \theta (t-0^{+})-\theta (t-\tau +0^{+})\right) $ in the cosine term
is inserted to weigh the even and odd configurations with opposite signs. In
order to estimate $ \Delta Z_{}^{aa^{\prime }}$ we use the mean field
approximation. In this approximation the grand partition functions of the
system are determined by the saddle point solutions, $\phi _s$, of the
effective functional $ S^{aa^{\prime }}$. The solutions $\phi _s$ correspond
to the equilibrium electrostatic potential of the plasma gas in the presence
of two external half charges located at points $0$ and $\tau $. In this way
one obtains that $\Delta Z_{}^{++}=\Delta Z_{}^{--}=0,$ while $\Delta
Z_{}^{+-}\left( \tau \right) =\Delta Z_{}^{-+}\left( \tau \right)
\rightarrow const$ when $\tau $ $\rightarrow \infty $. These results are
rather natural. The cancellation of $Z_e^{++}$ with $Z_o^{++}$ occurs
because the gas configurations with even and odd numbers of charges inside
the interval $\left( 0,\tau \right)$ are equally far away from the optimal
configuration. The latter should have inside the interval a charge equal to
$-\frac 12$ to screen the two external half charges of the same sign.
(Technically that cancellation occurs between contributions of different
saddle point solutions in the vicinity of consecutive minima of the cosine.
The existence of a manifold of minima reflects in a formal way the discreet
nature of the charges in the gas.) When the external half charges have
opposite signs the optimal configuration has an even number of charges in
the interval $\left( 0,\tau \right)$, and such a cancellation does not
occur. The value of the action $S^{+-}$ at the optimal configuration $\phi
_s$ determines the screened interaction between the two external charges.
For $\tau $ exceeding the screening radius $\tau _{scr}$ the bare
logarithmic interaction between the external charges is screened--off, and
therefore $\Delta Z_{}^{+-}\left( \tau \right) $ has a finite limit at large
$\tau $. In the mean field approximation $\Delta Z_{}^{+-}\left( \tau
\right) /Z$ $\equiv \Delta _g\sim \left( \eta /v_F\tau _{scr}\right)
^{g/2}$,where $\tau _{scr}= \frac{\eta g}{v_F}\left( \frac{g\delta _{-}}\pi
\right) ^{-1/\left( 1-g\right) }$.

Substituting these results into Eq.~(\ref{eq:gcg}) yields in the
asymptotic region $\tau \gg \tau_{scr}$
\begin{equation}
G(\tau ;0,0)\sim \frac 1\eta \exp \left( -\frac 12g^{-1}{\cal D}_{\widetilde{
\phi }}\right) \sim \frac 1\eta \left( \frac \eta {v_F\tau }\right)
^{1/2g}\Delta_g.  \label{eq:gtwbs}
\end{equation}
This result implies that the tunneling DOS, $\varrho (\epsilon ,0)$,
diverges in the infrared limit for a moderately repulsive electron--electron
interaction, $ 1/2<g<1:$ $\quad $
\begin{mathletters}
\begin{equation}
\varrho (\epsilon ,0)\sim \frac 1{\epsilon \eta }\left( \epsilon \eta
/v_F\right) _{}^{1/2g}\Delta _g\qquad \epsilon \ll \tau _{scr}^{-1}\quad
\label{eq:tdosa}
\end{equation}
(note that both $\tau _{scr}^{-1}$ and $\Delta _g$ go to
zero when $g\rightarrow 1$).

Up to now the electron--electron interaction was considered a
short--range one. However, for quantum wires the long--range character of
the Coulomb interaction may be important. The strength of the interaction
depends on the particular electrostatics of the sample. To include the
spatial dependence of the interaction amplitude, one should
substitute the combination $g^{-1}{\cal D}_{
\widetilde{\phi }}$ by $\int_0^{\eta ^{-1}}dpg(p)^{-1}\left(
1-e^{ipv_Ft}\right) /p$ in Eq.~(\ref{eq:gtwbs}); here
$g(p)=\sqrt{\frac{1-\gamma (p)}{1+\gamma (p)}}$,
 $\gamma (p)=\frac{V(p)}{\left( 2\pi v_F+V(p)\right) }$ and $V(p)$ is the
Fourier transform of the electron--electron interaction. For the Coulomb
interaction $V(p)=2e^2/\kappa \log \left( 1/\left| p\right| w\right) ,$
where $\kappa $ is the dielectric constant and $w$ is the width of the wire.
In this case in the asymptotic region $\tau \gg \tau _{scr}$ one gets that $
G(\tau ;0,0)$ $\sim \exp \left[ -\frac 12\frac 1{3\chi }\left( 1+2\chi \log
\frac{\tau v_F}\eta \right) ^{3/2}\right] ,$ where $\chi =e^2/\kappa \pi
\hbar v_F$. Inserting this expression into Eq.~(\ref{eq:tDOS}) yields that $
\varrho (\epsilon ,0)$ is nonmonotonic if $\tau
_{scr}^{-1}>\epsilon^{*}\sim\frac \eta {v_F}\exp \left( -\frac 32\chi
^{-1}\right) $.
 As $\epsilon $ decreases the DOS increases when $\tau _{scr}^{-1}>\epsilon
>\epsilon ^{*}$, and for $\epsilon ^{*}>\epsilon $ the DOS starts to vanish.
Since $\tau _{scr}$ is determined by $\delta _{-}$, while $\epsilon ^{*}$ is
not, the situation when $\tau _{scr}^{-1}$ is smaller than $\epsilon ^{*}$
is possible.

When the spin degrees of freedom of the conduction electrons are included,
the above considerations do not change essentially. The
calculation of the DOS can be reduced to the calculation of correlation
functions in a Coulomb gas. Due to the spin, the charge plasma contains two
types of particles. The latter aspect does not alter the physics
of screening. There are two fields $\phi _\rho $ and $\phi _\sigma $ (and
correspondingly two dual fields $\widetilde{\phi }_\rho $ and $\widetilde{
\phi }_\sigma $) that describe charge and spin density modes. The Green
function contains the factor $\exp -\frac 14\left( g_\rho ^{-1}{\cal D}_{
\widetilde{\phi }_\rho }+g_\sigma ^{-1}{\cal D}_{\widetilde{\phi }_\sigma
}\right) ,$ where $g_\sigma ^{}=1$ and $g_\rho ^{}=\sqrt{\frac{1-\gamma
_\rho }{1+\gamma _\rho }},$ $\gamma _\rho =\frac V{\left( \pi v_F+V\right) }=
\frac{2\gamma }{1+\gamma }.$ This factor is not influenced by the
screening and this yields

\begin{equation}
\varrho (\epsilon ,0)\sim \frac 1\epsilon \left( \epsilon \eta /v_F\right)
_{}^{1/4g_\rho +1/4}\Delta _g\qquad \epsilon \ll \tau _{scr}^{-1}\quad .\quad
\label{eq:tdosb}
\end{equation}
Therefore, in the spin case the tunneling DOS diverges when $1/3<g_\rho <1$.

Let us discuss now why the forward scattering does not influence the DOS. In
terms of the $\phi$--field operators the forward scattering can be written
as $\frac{\beta \delta _{+}}\pi \left. \frac{d\phi }{dx}\right|_{x=0}$,
where $\delta_{+}=U(k=0)/v_F$ and the amplitude $U(k=0)$ is the Fourier
component of the scattering potential with $k=0$.
   For the linearized spectrum the forward scattering can be absorbed in a
phase factor which finally disappears. To see this in a formal way, one can
apply the canonical transformation ${\cal U}=\exp \left( i\frac{ \delta
_{+}}{2\pi }g\beta \widetilde{\phi }\left( x=0\right) \right) $ which
removes the forward scattering term from the Hamiltonian, but produces a
phase factor in the backward scattering term and in the electron operators $
\psi _{R(L)}\left( x=0\right) $. However, due to the charge neutrality of
the Coulomb gas (including the external half charges) these phase factors
cancel each other out.

In a two--dimensional electron gas under the conditions of the quantum Hall
effect (QHE) the edge excitations are described by TL--like theories\cite
{FQHE:Wentot}. In the case of the integer QHE it is the interedge
interaction that leads to the TL--liquid behavior\cite{QHE:Oreg95}. The
above treatment of the backward scattering is not altered considerably.
However, since the particles which are moving on opposite edges are
spatially separated the amplitudes which describe the interedge ($ V_{er}$)
and the intraedge ($V_{ra}$) electron--electron interactions are not equal.
For that reason the expression for $\gamma $ should be modified. For the
symmetrical case, when the velocities of the excitations moving in opposite
directions are the same, $\gamma =\frac{V_{er}}{2\pi v_F+V_{ra}}$. Since the
electron liquid in the case of a fully occupied Landau level is
incompressible, it cannot screen off the long--range Coulomb interaction
between the edge electrons. In oreder to consider this effect the dependence
of $V_{er}(p)$ and $V_{ra}(p)$ on the momentum should be included as
discussed above.

Because of the nontrivial character of the electron liquid in the fractional
QHE state, electrons close to the edges exhibit an abnormal TL--type
behavior even in the absence of interedge interaction\cite {FQHE:Wentot}.
The backward scattering term describes now the scattering of fractionally
charged quasiparticles from one edge to the other. To find the asymptotic
behavior of the Green function of edge electrons for a filling factor $\nu
=1/n$, where $n$ is an odd integer, one should replace $g$ with $\nu $ in
expression (\ref{eq:gtwbs} ). This leads to $\varrho (\epsilon ,0)\sim
\epsilon ^{\frac n2-1}$ near the backward scattering center. This is a
considerable enhancement compared to $\varrho (\epsilon ,0)\sim \epsilon
^{n-1}$ in the absence of the backward scattering. When the interedge
electron--electron interaction is relevant, one gets $ \varrho (\epsilon
)\sim \epsilon ^{\frac 12\left( \nu g_\nu \right) ^{-1}-1}$ where $g_\nu
=\sqrt{\frac{1-\gamma _\nu }{1+\gamma _\nu }}$, $\gamma _\nu = \frac{\nu
V_{er}}{2\pi v_F+\nu V_{ra}}$. The long--range Coulomb interaction can be
treated as in the case of the 1--d quantum wire.

Let us discuss now the mechanism of the enhancement of the DOS. Following
the interpretation of the low energy physics of the backward scattering in a
1--d wire as a weak link junction, one would not expect an enhancement but a
vanishing of the DOS like $\epsilon ^{g^{-1}-1}$ for $g<1$
\cite{EE1D:Kane92a}. To understand the enhancement, note that the bosonic
representation (\ref {eq:f-b}) makes evident the affinity of the DOS with
the ''Debye--Waller factor'' of the $\phi$ mode. As a result of the backward
scattering together with the repulsion between electrons, the propagator of
small oscillations of the field $\phi \left( t, x=0 \right) $ acquires a
mass and becomes $\left( \left| \omega _n\right| +m\right) ^{-1}$, where
$\omega _n$ is a Matsubara frequency. The zero mode oscillations of the
$\phi $ mode become less effective and the Debye--Waller factor does not
vanish. Due to such pinning of the $\phi$ mode, the amplitude of an electron
created at the location of the backward scattering center, falls down slower
than in the case of free electrons (see Eq.~(\ref{eq:gtwbs})). Thus, because
of the multiple backward scattering, the escape rate of an electron from the
defect center slows down. The enhancement of the DOS is a consequence of
this effect. In the study of the Fermi edge singularity in the TL--liquid it
has been concluded \cite{XR:Oreg95} that the infrared physics of the
backward scattering problem resembles the physics of a Kondo resonance. We
believe that the enhancement of the tunneling DOS is a reminiscent of a
resonance of a similar type. We emphasize, however, that the treatment above
is not related directly to the analysis of the transport properties of
TL--liquids.

In summary, we have calculated the tunneling DOS at the location of a
backward scatterer in a 1--d quantum wire and for edge state electrons under
the conditions of the QHE. A singular enhancement of the DOS was obtained.
The enhancement of the DOS in the TL--liquid may be observed not only when
the backward scattering is due to an internal defect. When the
counterelectrode in a tunneling experiment has the shape of a sharp tip,
then the tip itself may cause a backward scattering of the conduction
electrons.

We would like to thank A.~Kamenev and D.~Orgad for numerous discussions.
A.F. is supported by the Barecha Fund Award. The work is supported by the
Israel Academy of Science, Grant No.~801/94--1.]
\end{mathletters}

\bibliographystyle{prsty}

\end{document}